\documentclass{IEEEtran}

\usepackage{cite}
\usepackage{amsmath,amssymb,amsfonts}
\usepackage{newtxtext,newtxmath}

\usepackage{graphicx}
\usepackage{textcomp}
\usepackage{multirow}
\usepackage{subfig}
\usepackage{booktabs}
\usepackage{array}
\usepackage{stfloats}

\usepackage{hyperref}
\hypersetup{
    colorlinks=true,
    linkcolor=black,
    citecolor=black,
    urlcolor=black
}

\usepackage{lineno}

\usepackage{xcolor}
\usepackage{tikz}
\usetikzlibrary{arrows.meta,positioning,shapes.geometric,fit}

\def\BibTeX{{\rm B\kern-.05em{\sc i\kern-.025em b}\kern-.08em
    T\kern-.1667em\lower.7ex\hbox{E}\kern-.125emX}}

\begin{document}
\title{Efficient and High-Accuracy Ray Tracing in Discretized Ionospheric Models}

\author{Qinglin Li,~Wen Liu,~Zhigang Zhang,~Fengjuan Sun,~Rong Chen,~Zhongxin Deng,\\
~and~Zhiqiang Yao,~\IEEEmembership{Senior Member,~IEEE}%
\thanks{This work was supported in part by the National Key Research and Development Program of China under Grant 2023YFA1009100, 
and in part by the Postgraduate Scientific Research Innovation Project of Xiangtan University under Grant XDCX2024Y187.}%
\thanks{Qinglin Li is with the School of Mathematics and Computational Science, Xiangtan University, Xiangtan 411105, China. 
Wen Liu, Rong Chen, Zhongxin Deng, and Zhiqiang Yao are with the School of Automation and Electronic Information, Xiangtan University, Xiangtan 411105, China (corresponding author: Wen Liu; e-mail: l\_wen9209@xtu.edu.cn).}%
\thanks{Zhigang Zhang is with the Naval University of Engineering, Wuhan 430010, China. Fengjuan Sun is with the China Research Institute of Radiowave Propagation, Qingdao 266107, China.}%
\thanks{\textcopyright~2026 IEEE. Personal use of this material is permitted.
Permission from IEEE must be obtained for all other uses, in any current
or future media, including reprinting/republishing this material for
advertising or promotional purposes, creating new collective works,
for resale or redistribution to servers or lists, or reuse of any
copyrighted component of this work in other works.
DOI: 10.1109/TAP.2026.3681396.}%
}

\maketitle

\begin{abstract}
High-frequency (HF) ray tracing in complex ionospheric media generally involves a fundamental trade-off between path accuracy and computational efficiency, which directly affects practical applications such as over-the-horizon radar, ionospheric monitoring, and HF skywave communication systems. 
This paper presents RTM-GD, a ray-tracing framework that combines Hamiltonian ray integration with a continuously differentiable Galerkin--Difference (GD) interpolation strategy. 
Under discretized ionospheric grid conditions, the electron density is reconstructed as a $C^1$-continuous function within each grid cell, yielding an everywhere differentiable electron-density field for stable numerical integration and improved propagation-path accuracy. 
Numerical simulations and validations using measured HF oblique sounding data are conducted under diverse conditions, including different ionospheric states, low- and high-elevation angles, multiple operating frequencies, and both ordinary (O) and extraordinary (X) wave modes. 
Results show that RTM-GD consistently achieves sub-kilometer RMSE in both group-path and ground-distance metrics and sub-0.01-degree azimuth deviation relative to Richardson extrapolation, while reducing computational time by 98\%. 
Compared with Catmull--Rom interpolation, RTM-GD reduces the RMSEs of ray parameters by approximately one order of magnitude with less than 4\% additional computational cost. 
Measured-data validation based on ionogram synthesis further shows that the mean relative group-path error remains within 7\%, confirming reliable reproduction of practical HF oblique propagation characteristics. 
Overall, RTM-GD provides an accurate and computationally efficient framework for HF ray tracing in discretized ionospheric environments.
\end{abstract}


\section{Introduction}
\label{sec:introduction}

\IEEEPARstart{H}{igh}-frequency (HF) applications such as over-the-horizon radar and shortwave source localization rely critically on ionospheric propagation. 
The spatial structure and dynamic variability of the ionosphere significantly influence signal trajectories, localization accuracy, 
and system reliability~\cite{budden1985radio, davies1990ionospheric}. Achieving high-precision HF propagation modeling demands 
accurate and efficient computation of wave propagation parameters. 
Recent studies have further highlighted the importance of accurate ionospheric propagation modeling in HF applications~\cite{Zhang2025SW_MUF,Wang2023TAP_MUF,Shi2024TAP_HFprop,Wang2022ASR_review}. 
However, the ionosphere presents significant modeling challenges due to its inherent inhomogeneity, anisotropy, dispersion, and temporal variability~\cite{kelley2009ionosphere}. 

To address this, three-dimensional ray tracing remains the predominant approach for modeling electromagnetic wave trajectories in the ionosphere. 
Hamiltonian ray tracing based on the Haselgrove ray differential equations is widely used as the theoretical foundation for ionospheric propagation modeling, 
in which ray trajectories are obtained by numerically integrating a canonical system of equations describing wave evolution 
in inhomogeneous and anisotropic media~\cite{haselgrove1955ray,jones1968threedimensional}.
IONOLAB-RAY discretizes the ionosphere into a three-dimensional spherical coordinate grid and assumes the electron density and related physical parameters to be constant 
within each grid cell. Ray propagation directions are then determined by applying Snell’s law at interfaces between adjacent cells, 
providing an efficient and practically oriented approach for fast shortwave ray tracing\cite{erdem2017ionolab}. 
In this work, we adopt the Hamiltonian ray tracing framework based on the Haselgrove formulation, which requires numerical integration of ray differential equations 
involving spatial derivatives of the electron density field. 

Modern ionospheric models typically represent the electron density $N_e$ 
as a discretized three-dimensional grid reconstructed from data assimilation, GNSS tomography, or empirical climatology~\cite{prol2024gnss, zhao2025regional}. 
However, such discretized representations do not directly provide smooth and reliable spatial gradients. As a result, numerical ray integration may suffer from 
inaccurate curvature evaluation, degraded trajectory fidelity, and potential instability, especially in regions where electron density varies rapidly. 
To address this issue, it is necessary to reconstruct an electron density field that is differentiable everywhere over the discretized ionospheric grid. One viable approach 
is to interpolate the electron density within each grid cell as a $C^1$-continuous function, thereby enabling stable and accurate numerical integration of the ray differential equations.

\subsection{Related Work}
To better understand the challenges of gradient reconstruction in gridded models, it is helpful to revisit 
how electron density and its derivatives have been represented in earlier ionospheric models. 
These approaches, though limited in flexibility, reveal fundamental limitations that 
have motivated the shift toward gridded representations supported by interpolation.

For example, \cite{coleman1998ray} employs a three-layer Chapman model 
with cubic splines for 2D grids with vertical gradients, though limited 
to two-dimensional scenarios. \cite{sokolov2016recent} implements a global 
3D tricubic spline model using assimilative ionospheric data, but without 
explicit error analysis or derivative validation. 
The PHaRLAP toolbox applies Lagrange interpolation within the IRI model~\cite{cervera2014modeling}, 
while the IONORT system adopts piecewise polynomial schemes~\cite{azzarone2012ionort}. 
Both approaches do not explicitly guarantee $C^1$ continuity across grid cell boundaries.
Additionally, the GPSII model integrates three-dimensional Catmull–Rom splines, reporting approximately 30\% runtime 
improvement without formal accuracy evaluation~\cite{nickisch2016feasibility}. 
Finally, a hybrid model using spherical harmonics and $\alpha$-Chapman 
functions is developed in~\cite{tsai2010three}, which suffers from basis 
mismatch and limited regional fitting performance.

Overall, in most existing studies, interpolation schemes are introduced mainly as auxiliary numerical tools. 
The resulting models often do not ensure $C^1$ continuity of the reconstructed electron density field and 
lack a systematic analysis of their mathematical and numerical properties. 
Moreover, the influence of interpolation strategies on the trade-off between 
ray tracing accuracy and computational efficiency has not been sufficiently investigated.

\subsection{Motivations and Contributions}
Under discretized ionospheric models, constructing an electron density reconstruction method that achieves both high accuracy and 
computational efficiency remains an open problem. This challenge is closely related to the long standing accuracy and efficiency trade-off in ray tracing.
Motivated by this challenge, this work investigates the mathematical properties of interpolation models and their influence 
on ray tracing performance from a mechanistic perspective, and validates the proposed approach through numerical simulations and experiments with measured data.
The main contributions of this work are summarized as follows:
\begin{itemize}
\item \textbf{Theoretical analysis of interpolation accuracy and numerical properties.} 
Theoretical analyses of interpolation accuracy and computational complexity are conducted to characterize 
the error behavior of interpolation models and to examine their impact on ray-tracing performance.
\item \textbf{A $C^1$-continuous electron density reconstruction based on Galerkin–Difference interpolation.} 
Galerkin–Difference (GD) interpolation is incorporated into the ray-tracing framework to construct an everywhere differentiable
electron density field. This enables stable numerical integration of the ray differential equations and provides analytical expressions 
for electron density gradients, thereby improving propagation-path accuracy while accelerating numerical ray integration.
\item \textbf{Systematic validation across diverse ionospheric propagation scenarios.}
Extensive numerical simulations and experimental validations are conducted under diverse conditions, 
including different ionospheric states, low- and high-elevation angles, multiple operating frequencies, 
and both ordinary (O) and extraordinary (X) wave modes. 
The results comprehensively demonstrate the proposed method’s accuracy, numerical stability, and computational efficiency.
\end{itemize}

The remainder of the paper is organized as follows. 
Section~\ref{sec:method} introduces the Hamiltonian formulation for 3D ray tracing and the proposed RTM-GD method, including its interpolation model construction and algorithmic implementation.
Section~\ref{sec:theory_analysis} provides theoretical analysis of interpolation accuracy and computational complexity.  
Section~\ref{sec:experiment} presents numerical simulations and experiments using measured HF data of RTM-GD for ionospheric reconstruction and HF propagation.
Section~\ref{sec:conclusion} concludes the paper and outlines future work.

\section{Methodology}
\label{sec:method}
This section presents the methodological framework of the proposed RTM-GD approach, 
including the Haselgrove-based ray-tracing formulation, the GD interpolation method for electron-density reconstruction, 
and the algorithmic implementation of RTM-GD.

\subsection{Hamiltonian Ray Tracing}
\label{sec:hamilton_gradient}

High-frequency (HF) wave propagation in an inhomogeneous ionosphere can be formulated as a Hamiltonian system in 
spherical coordinates $(r,\theta,\phi)$~\cite{haselgrove1955ray,jones1968threedimensional}. 
Let $(k_r,k_\theta,k_\phi)$ denote the wave–vector components and $\omega$ the angular frequency. 
The ray trajectory evolves according to
\begin{subequations}
\begin{align}
\frac{dr}{dP'} &=
-\frac{1}{c}\left( \frac{\partial H}{\partial k_r}\middle/
\frac{\partial H}{\partial \omega}\right), \\
\frac{d\theta}{dP'} &=
-\frac{1}{rc}\left( \frac{\partial H}{\partial k_\theta}\middle/
\frac{\partial H}{\partial \omega}\right), \\
\frac{d\phi}{dP'} &=
-\frac{1}{rc\sin\theta}\left( \frac{\partial H}{\partial k_\phi}\middle/
\frac{\partial H}{\partial \omega}\right), \\
\frac{dk_r}{dP'} &=
\frac{1}{c}\left( \frac{\partial H}{\partial r}\middle/
\frac{\partial H}{\partial \omega}\right)
+ k_\theta\frac{d\theta}{dP'} + k_\phi\frac{d\phi}{dP'}, \\
\frac{dk_\theta}{dP'} &=
\frac{1}{r}\!\left[
\frac{1}{c}\!\left(\frac{\partial H}{\partial \theta}\middle/
\frac{\partial H}{\partial \omega}\right)
- k_\theta\frac{dr}{dP'}
+ k_\phi r\cos\theta \frac{d\phi}{dP'}
\right], \\
\frac{dk_\phi}{dP'} &=
\frac{1}{r\sin\theta}\!\left[
\frac{1}{c}\!\left(\frac{\partial H}{\partial \phi}\middle/
\frac{\partial H}{\partial \omega}\right)
- k_\phi\sin\theta\frac{dr}{dP'}
- k_\phi r\cos\theta\frac{d\theta}{dP'}
\right].
\end{align}
\end{subequations}
The Hamiltonian for a cold, magnetized plasma is
\begin{equation}
H = \frac{1}{2}\left[\frac{c^2}{\omega^2}
(k_r^2+k_\theta^2+k_\phi^2) - \mu^2 \right],
\label{eq:hamiltonian_new}
\end{equation}
where the refractive index $\mu$ is determined by the Appleton-Hartree relation:
\begin{equation}
\mu^2 = 1 - 
\frac{2X(1-X)}{2(1-X) - Y_T^2
\pm\sqrt{Y_T^4 + 4Y_L^2(1-X)^2}}.
\label{eq:appleton_new}
\end{equation}
Here, $X=f_N^2/f^2$ and $Y=f_H/f$ denote the plasma-frequency and gyrofrequency ratios, 
with the signs corresponding to the ordinary (O) and extraordinary (X) modes. 
Since $\mu$ depends on both the electron density $N_e$ and the geomagnetic field $\mathbf{B}$, 
the spatial derivatives $\nabla N_e$ directly affect $\nabla H$ and therefore the curvature, smoothness, and stability of the ray path.

Modern ionospheric models provide $N_e$ on structured 3D grids, so both $N_e$ and its gradients must be reconstructed numerically at every integration step. 
For a single ray, the total computation time can be written as
\begin{equation}
T = T_{\text{IonoGrad}} + T_{\text{MagGrad}} + T_{\text{ODE}},
\end{equation}
where the ionospheric contribution $T_{\text{IonoGrad}}$ often dominates the runtime, 
since $\nabla N_e$ is typically obtained via numerical differentiation on the grid, 
which is considerably more expensive than evaluating analytical geomagnetic-field gradients or advancing the ray state with the ODE solver.

The accumulated trajectory error admits the decomposition
\begin{equation}
\epsilon_{\text{total}}^2 =
\epsilon_{\text{IonoModel}}^2 +
\epsilon_{\text{MagModel}}^2 +
\epsilon_{\text{OdeSolver}}^2 +
\epsilon_{\text{Interp}}^2,
\end{equation}
where $\epsilon_{\text{Interp}}$ denotes the error introduced by interpolating $N_e$ and its gradient. 
Discontinuities or low‐order approximations in $\nabla N_e$ can cause nonphysical refraction, 
oscillatory paths, and numerical instability.
Therefore, a smooth, high‐order, and $C^{1}$‐continuous reconstruction of the electron density 
is essential for robust HF ray tracing. 
This requirement motivates the development of the interpolation scheme introduced in the next subsection.

\subsection{Interpolation Models for Electron-Density Reconstruction}
\label{sec:interp_models}
Ray tracing in a discretized ionospheric model requires evaluating
the electron density $N_e$ and its spatial gradients $\nabla N_e$
at arbitrary positions within a three-dimensional grid.
Since the numerical integration of the ray equations is sensitive
to the smoothness of the reconstructed field, appropriate interpolation
schemes are essential.
In this work, we consider two local interpolation methods:
the classical Catmull--Rom (CR) spline and the proposed
Galerkin--Difference (GD) interpolant.
Both methods operate on structured grids and are extended to
three dimensions via tensor products of their one-dimensional
basis functions.

\subsubsection{Catmull--Rom Spline Interpolation}

Catmull--Rom spline~\cite{catmull1974class} is a $C^1$-continuous local cubic Hermite
interpolation method. It estimates first-order derivatives using central differences
and constructs smooth interpolants from four adjacent grid points.

On a one-dimensional grid \(\{x_j\}\), the interpolated function is given by
\begin{equation}
\tilde{u}(x) = \sum_{k=-1}^2 u_{j+k} w_k(t),
\label{eq:1dinterp}
\end{equation}
where
\[
t = \frac{x - x_j}{x_{j+1} - x_j}, \quad x_j \le x < x_{j+1},
\]
and \( w_k(t) \) are the standard Catmull--Rom basis functions,
\begin{equation}
\begin{bmatrix}
w_{-1}(t) \\[4pt]
w_0(t) \\[4pt]
w_1(t) \\[4pt]
w_2(t)
\end{bmatrix}
= \frac{1}{2}
\begin{bmatrix}
 -t^3 + 2t^2 - t \\
 3t^3 - 5t^2 + 2 \\
 -3t^3 + 4t^2 + t \\
 t^3 - t^2
\end{bmatrix}.
\label{eq:cr_basis}
\end{equation}

The three-dimensional CR interpolation is constructed via tensor products of
the one-dimensional bases,
\begin{equation}
\begin{split}
\tilde{u}(x,y,z) =
\sum_{\alpha=-1}^2 \sum_{\beta=-1}^2 \sum_{\gamma=-1}^2
u_{i+\alpha,j+\beta,k+\gamma}
\, w_\alpha^x(\xi_x) w_\beta^y(\xi_y) w_\gamma^z(\xi_z),
\end{split}
\label{eq:3dinterp_cr}
\end{equation}
where the local coordinates are defined as
\begin{equation}
\begin{split}
\xi_x &= \frac{x - x_i}{x_{i+1} - x_i}, \\
\xi_y &= \frac{y - y_j}{y_{j+1} - y_j}, \\
\xi_z &= \frac{z - z_k}{z_{k+1} - z_k},
\end{split}
\label{eq:local_coords_cr}
\end{equation}
with \(\xi_x, \xi_y, \xi_z \in [0,1)\).

\subsubsection{Galerkin--Difference Interpolation}\label{sec:gd model}

The Galerkin--Difference (GD) interpolation constructs a locally supported,
$C^1$-continuous piecewise polynomial approximation on a uniform structured grid,
originally developed for problems requiring smooth function values and
first-order derivatives~\cite{jacangelo2020galerkin}.
Such properties are particularly important for ionospheric ray tracing,
where stable numerical integration of the Haselgrove ray equations
relies on continuous electron-density gradients.

On each cell $x_j < x \le x_{j+1}$ with uniform spacing $h=x_{j+1}-x_j$,
the interpolant is defined as
\begin{equation}
\tilde{u}(x) = \tilde{u}_j(x),
\end{equation}
where $\tilde{u}_j(x)$ is a degree-$(p+2)$ polynomial constructed from
$p+1$ neighboring nodal values.
Introducing the normalized local coordinate
\begin{equation}
\xi = \frac{x - x_j}{h}, \quad \xi \in (0,1],
\end{equation}
the GD interpolant can be written in Hermite form as
\begin{equation}
\tilde{u}_j(x)
= \sum_{\alpha} H_{\alpha,0}^{(p,1)}(\xi)\, u_{j+\alpha}
+ \sum_{\alpha=0}^{1} H_{\alpha,1}^{(p,1)}(\xi)\, h\, D^{(1,p)} u_{j+\alpha},
\label{eq:interpolant}
\end{equation}
where the first term interpolates nodal values and the second term enforces
first-derivative continuity at cell interfaces through a central
finite-difference operator $D^{(1,p)}$.
This construction ensures global $C^1$ continuity of $\tilde{u}(x)$ across the grid.
The explicit Hermite basis functions and finite-difference coefficients
are omitted here for brevity and can be found in~\cite{jacangelo2020galerkin}.

For completeness, the corresponding one-dimensional global GD basis function can be written as
\begin{equation}
\begin{aligned}
\psi_k^{(p,1)}(x)
={}& H_{-k,0}^{(p,1)}(\xi-k)
+ \eta_{-k}^{(1,p)} H_{0,1}^{(p,1)}(\xi-k) \\
&+ \eta_{-k-1}^{(1,p)} H_{1,1}^{(p,1)}(\xi-k),
\qquad k \le \xi \le k+1,
\end{aligned}
\label{eq:psi_case1}
\end{equation}
and $\psi_k^{(p,1)}(x)=0$ otherwise.

For a three-dimensional uniform grid, the GD basis functions are constructed
via tensor products of the one-dimensional bases.
Let
\begin{equation}
\xi_x = \frac{x-x_i}{h_x}, \quad
\xi_y = \frac{y-y_j}{h_y}, \quad
\xi_z = \frac{z-z_k}{h_z},
\end{equation}
then the interpolated electron density is given by
\begin{equation}
\begin{aligned}
\tilde{u}(x,y,z)
={}& \sum_{\alpha=-q+1}^{q}
\sum_{\beta=-q+1}^{q}
\sum_{\gamma=-q+1}^{q}
u_{i+\alpha,j+\beta,k+\gamma} \\
&\times \psi_\alpha^{(p,1)}(\xi_x)
\psi_\beta^{(p,1)}(\xi_y)
\psi_\gamma^{(p,1)}(\xi_z),
\end{aligned}
\label{eq:3dinterp_gd}
\end{equation}
where $h_x$, $h_y$, and $h_z$ denote the uniform grid spacings in each dimension.
The resulting GD basis functions have compact support over a local stencil
of width $2q$ in each spatial dimension, and each interpolation involves
only a finite number of neighboring grid nodes.

The GD scheme admits odd polynomial orders $p=1,3,5,\dots$,
providing a trade-off between interpolation accuracy, numerical stability,
and computational cost.
Based on the theoretical analysis in Section~\ref{sec:theory_analysis}
and the grid-sensitivity experiments in Section~\ref{sec:GridSensitivity},
we adopt $p=3$ as the default configuration in this work.
This choice yields sufficiently smooth $C^1$-continuous gradients for stable
ray integration while avoiding the excessive computational overhead
associated with higher-order schemes.

For completeness, when $p=3$ the GD interpolation employs a five-point stencil
($q=2$) and constructs piecewise quintic polynomials.
The explicit basis functions and finite-difference weights are provided
in Appendix~\ref{appendix:gd_p3}.

\subsection{Algorithm Implementation}
\label{sec:algorithm}
\begin{figure}[!t]
   \centering
   \includegraphics[width=0.95\linewidth]{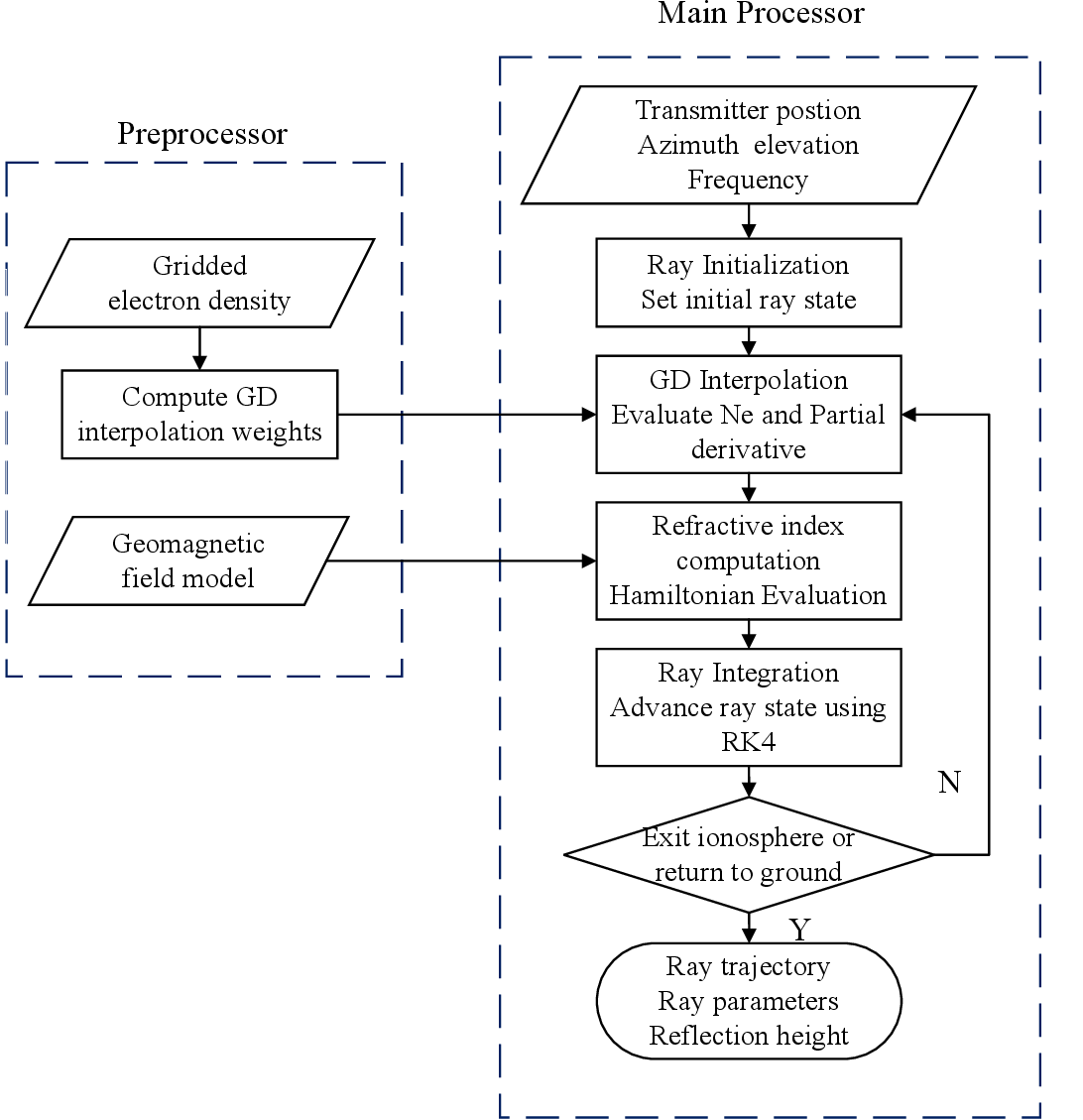}
   \caption{Flowchart of the proposed RTM-GD method.}
   \label{fig:flowchart}
\end{figure}

The ray-tracing algorithm developed in this study, referred to as RTM-GD, is designed for high-frequency (HF) wave 
ray tracing in a three-dimensional discretized ionospheric model. 
The algorithm is composed of two functional modules, 
namely a preprocessing module and a main processing module. 
A flowchart illustrating the overall computational workflow 
and the interaction between these two modules is shown in Fig.~\ref{fig:flowchart}

In the preprocessing stage, the ionosphere is represented by a gridded electron density distribution 
defined over a prescribed geographic region. 
The gridded electron density serves as the primary physical input to the algorithm.

Based on this discretized electron density field, 
the GD interpolation weights are computed 
for each grid cell. These weights enable continuous and differentiable evaluation 
of the electron density and its spatial partial derivatives 
during ray propagation. Since they are independent of individual ray trajectories, 
the GD interpolation weights are computed once in the preprocessing stage 
and reused throughout the ray-tracing procedure.
In addition, a geomagnetic field model is incorporated during preprocessing to 
provide the geomagnetic field representation required for refractive index evaluation.

In the main processing stage, the user specifies the transmitter position, 
launch azimuth, elevation angle, and operating frequency as input parameters. 
Based on these inputs, the ray is initialized.
At each propagation step, the GD interpolation module is invoked to evaluate 
the local electron density and its spatial partial derivatives at the current ray position. 
These quantities, together with the geomagnetic field information provided 
by the preprocessing module, are used to compute the refractive index 
and the corresponding Hamiltonian governing ray propagation.
The ray state is then advanced by numerically integrating the Hamiltonian ray equations 
using a fourth-order Runge–Kutta (RK4) scheme. This integration process 
is repeated iteratively, until the ray either exits the ionospheric region 
or returns to the ground. Upon termination, the algorithm outputs the complete 
three-dimensional ray trajectory, together with key ray parameters 
including group path length, ground distance, azimuth deviation, and the reflection height.

All simulations were conducted on a desktop computer equipped with a
12th-generation Intel Core i7-12700 processor (12 cores, 20 threads, 2.10~GHz)
and 16~GB RAM, running Windows~11. The RTM-GD framework was implemented in C++
using Visual Studio~2022. In the present study, the IRI2020 and IGRF13 models were
used as representative data sources to generate gridded electron density and
geomagnetic field inputs; however, the proposed RTM-GD algorithm is input-agnostic
and is directly applicable to electron density grids produced by arbitrary
discretized ionospheric models.

\section{Interpolation Error and Complexity Analysis}
\label{sec:theory_analysis}
This section analyzes the interpolation schemes used in RTM-GD from the perspectives of approximation quality and computational cost. 
We first present error estimates for the CR and GD reconstructions, and then compare their evaluation complexity in one and three dimensions. 
These results help clarify the tradeoff between numerical accuracy and efficiency, and provide theoretical support for the use of GD interpolation in high-fidelity ionospheric ray tracing.
\subsection{Error Analysis}\label{sec:errorbound}
This interpolation error analysis assumes a structured and uniform grid, consistent with the discretized ionospheric models commonly used in ray tracing. The high-order accuracy and gradient continuity of GD interpolation depend on the regularity of the grid, and the findings are particularly relevant for grid-based electron density reconstructions.
This subsection analyzes the theoretical interpolation errors of CR spline and GD spline models in both 1D and 3D settings.
\subsubsection{Catmull-Rom Spline}
For a function \( u(x) \in C^4 \), the interpolation error satisfies
\begin{equation}
\| u(x) - \tilde{u}(x) \| \leq C h^4 \max_{x \in [x_j, x_{j+1}]} \left| u^{(4)}(x) \right|,
\end{equation}
where \( h \) is the uniform grid spacing and \( C \) is a constant independent of \( h \) and \( u \).

In three dimensions, for \( u(x,y,z) \in C^4(\Omega) \), the error bound generalizes to
\begin{equation}
\begin{split}
\| u - \tilde{u} \| \leq {} & C \bigl( h_x^4 + h_y^4 + h_z^4 \bigr) \\
& \times \max \left\{ \|\partial_x^4 u\|_\infty, \|\partial_y^4 u\|_\infty, \|\partial_z^4 u\|_\infty \right\}.
\end{split}
\end{equation}
where \( h_x, h_y, h_z \) denote the grid spacings in the \( x \), \( y \), and \( z \) directions, respectively.

Despite providing \( C^1 \) continuity and fourth-order accuracy, the CR spline interpolation has limitations on nonuniform grids and near boundaries, and exhibits reduced smoothness in regions with high gradients.

\subsubsection{Galerkin-Difference Interpolation}
We next examine the approximation error of the GD interpolation scheme. 
Let \(u \in C^{p}(\Omega)\) be a sufficiently smooth scalar field defined on a structured Cartesian grid, and let \(\tilde{u}\) denote its GD reconstruction. 
For any point \((x,y,z)\) in the interpolation domain, the local interpolation error is defined by
\begin{equation}
E^{(p)}(x,y,z) = u(x,y,z) - \tilde{u}(x,y,z).
\end{equation}

Using the Peano-kernel representation~\cite{davis1984methods}, the one-dimensional interpolation remainder can be expressed in integral form as
\begin{equation}
E^{(p)}(x) = \int_{x_0}^{x_1} \frac{d^p u}{dt^p}(t) \cdot K^{(p)}(t,x) \, dt,
\end{equation}
where \(K^{(p)}(t,x)\) denotes the Peano kernel associated with the interpolation operator and its local stencil. 
For the tensor-product GD construction in three dimensions, the corresponding remainder can be written as a triple integral over the interpolation domain:
\begin{equation}
\begin{split}
E^{(p)}(x,y,z) = \iiint_\Omega 
& \frac{\partial^p u}{\partial t_1^p \partial t_2^p \partial t_3^p}(t_1,t_2,t_3) \\
& \cdot K^{(p)}(\mathbf{t}; \mathbf{x}) \, dt_1 dt_2 dt_3,
\end{split}
\label{eq:gd_peano}
\end{equation}
where $\mathbf{t} = (t_1, t_2, t_3)$ and $\mathbf{x} = (x, y, z)$, and $K^{(p)}$ is the 3D tensor-product Peano kernel constructed from the GD basis.

Because the GD reconstruction is built from \(C^{1}\) Hermite-type basis functions of degree \(p+2\), and because the interface derivatives are enforced through central-difference operators with accuracy \(\mathcal{O}(h^{p-1})\), the resulting interpolation error is controlled by the \(p\)-th derivatives of \(u\) together with the local mesh sizes. 
This leads to the following conservative estimate:
\begin{equation}
\begin{split}
|u(x,y,z) - \tilde{u}(x,y,z)| \le {} & C_p \cdot (h_x^p + h_y^p + h_z^p) \\
& \times \max_{\Omega} \left\{
\left|\frac{\partial^p u}{\partial x^p}\right|,
\left|\frac{\partial^p u}{\partial y^p}\right|,
\left|\frac{\partial^p u}{\partial z^p}\right|
\right\},
\end{split}
\label{eq:gd_interp_error_bound}
\end{equation}
where $h_x$, $h_y$, and $h_z$ denote grid spacings in each direction, and $C_p$ is a constant depending on $p$ and the basis support.

A similar argument can be applied to the reconstructed gradient. 
By differentiating the GD interpolant and combining the truncation error of the derivative stencil, one obtains the following estimate for the gradient reconstruction error:
\begin{equation}
\begin{split}
|\nabla u - \nabla \tilde{u}| \le {} & C_p' \cdot (h_x^{p-1} + h_y^{p-1} + h_z^{p-1}) \\
& \times \max_{\Omega} \left\{
\left|\frac{\partial^p u}{\partial x^p}\right|,
\left|\frac{\partial^p u}{\partial y^p}\right|,
\left|\frac{\partial^p u}{\partial z^p}\right|
\right\},
\end{split}
\label{eq:gd_grad_error_bound}
\end{equation}
where $C_p'$ is another constant depending on the interpolation stencil. 
This estimate indicates that GD interpolation preserves high-order consistency not only for function values 
but also for first-order spatial derivatives, which is particularly important for gradient-sensitive Hamiltonian ray integration.

\subsection{Computational Complexity Analysis}\label{sec:complexity_analysis}
We now analyze the computational complexity of the CR and GD interpolation schemes
in terms of their per-point evaluation cost,
to assess their suitability for efficient electron-density reconstruction
in grid-based ray-tracing applications.

\subsubsection{Catmull--Rom Spline}

The computational complexity of Catmull--Rom (CR) interpolation
is primarily determined by its compact local support.
In one dimension, as shown in Eq.~\eqref{eq:1dinterp},
the interpolated value at any point depends only on four neighboring
grid nodes and the evaluation of the corresponding basis functions.
Since the basis functions \(w_k(t)\) in Eq.~\eqref{eq:cr_basis}
are fixed cubic polynomials, the per-point interpolation cost
consists of a small number of polynomial evaluations and weighted
summations, resulting in a constant-time operation, denoted as \(O(1)\).

In three dimensions, the tensor-product formulation in
Eq.~\eqref{eq:3dinterp_cr} requires a weighted summation over
\(4 \times 4 \times 4 = 64\) neighboring grid nodes.
Although the computational workload increases compared to the
one-dimensional case, the stencil size remains fixed.
Consequently, the per-point evaluation cost of three-dimensional
CR interpolation also exhibits constant-time complexity \(O(1)\).

In summary, owing to its compact support and low-degree polynomial basis,
CR interpolation offers low and predictable computational cost per
interpolation point.
These properties make it well suited for efficient interpolation
in large-scale ray-tracing simulations and amenable to parallel
implementation.

\subsubsection{Galerkin--Difference Interpolation}
\label{sec:gd_complexity}
The computational cost of GD interpolation is mainly determined by the stencil width \(2q\), whereas the polynomial order \(p\) primarily affects basis evaluation rather than the stencil extent. 
In one dimension, evaluating \(\tilde{u}_{j}(x)\) within a cell requires contributions from \(2q\) nearby nodal values together 
with derivative-related terms generated by the central-difference coefficients. 
As a result, the arithmetic cost grows linearly with the stencil width.

Because the GD basis functions have compact local support, a one-dimensional interpolation query remains a local operation with \(\mathcal{O}(q)\) complexity. 
In three dimensions, the tensor-product construction enlarges the support to \((2q)^3\) grid nodes, so the per-point evaluation cost scales as \(\mathcal{O}(q^3)\). 
Even in this case, the computation is still entirely local and therefore suitable for parallel implementation.

Compared with CR interpolation, GD interpolation introduces a moderate additional per-query cost. 
However, this overhead is compensated by the smoother gradient reconstruction and improved derivative fidelity, 
which are essential for stable and accurate ray tracing in discretized ionospheric models.

\section{Experiment and Discussion}\label{sec:experiment}
In this section, we present a comprehensive experimental evaluation 
to systematically assess the applicability and performance of the 
RTM-GD method in ionospheric reconstruction and HF wave propagation. 
The experiments are designed to evaluate numerical accuracy, path 
stability, and the capability of reproducing actual radio wave propagation 
features. First, we conduct a grid-resolution sensitivity experiment 
to verify the convergence behavior and directional response characteristics 
of our interpolation method across various spatial scales. Subsequently, 
an elevation-angle scanning experiment covering from the E-layer to the 
F-layer assesses parameter accuracy in ray tracing and computational 
efficiency of the RTM-GD. We then focus specifically on high-elevation-angle 
propagation in the F2 layer, to evaluate trajectory stability and error control. Finally, 
we validate the model's ability to accurately replicate real propagation 
behavior by comparing synthetic oblique ionograms with actual observational 
data. 

\subsection{Sensitivity to Grid Resolution}
\label{sec:GridSensitivity}
To verify the theoretical error order $\mathcal{O}(h^{p+1})$ of the GD interpolation derived in Section~\ref{sec:errorbound}, 
we conducted a grid-resolution sensitivity experiment. Electron density data used were generated using the IRI-2020 model 
within the altitude range from 65\,km to 500\,km, and latitudes and longitudes from $5^\circ$N--$28^\circ$N and $113^\circ$E--$146^\circ$E. 
To evaluate interpolation performance under representative ionospheric conditions, 
six grid configurations (V1-V6) were designed to reflect three typical scenarios: 
V1 represents a coarse-resolution baseline, V2-V3 emphasize vertical refinement to capture strong 
vertical gradients in the lower ionosphere and transition region, and V4-V6 focus on horizontal 
refinement to assess sensitivity to latitudinal and longitudinal variations.
The accuracy of our GD interpolation (with orders $p=3$ and $p=5$) was compared against traditional CR interpolation by analyzing root-mean-square errors (RMSE).
\begin{equation}
    \mathrm{Ne}_{\mathrm{RMSE}} = \sqrt{\frac{1}{N} \sum_{i=1}^N \left( \mathrm{Ne}_{\text{interp},i} - \mathrm{Ne}_{\text{IRI},i} \right)^2}
    \label{eq:ne_rmse}
\end{equation}

Table~\ref{interp_rmse} shows that the relative performance of GD and CR interpolation varies markedly across altitude ranges, 
which reflects the strong vertical inhomogeneity of the ionosphere. 
In the lower ionosphere, the interpolation error is governed mainly by the vertical grid spacing. 
For coarse grids, the higher-order GD scheme may exhibit mild oscillatory behavior, 
but this effect is rapidly suppressed as the vertical resolution is refined, 
after which the expected convergence trend becomes evident.

In the transition region, GD interpolation maintains stable accuracy across the tested grid settings, 
indicating good robustness in regions with strong gradients. 
By contrast, CR interpolation is more sensitive to horizontal resolution changes, especially in latitude. 
At higher altitudes, where the electron-density profile becomes smoother, 
GD continues to benefit from mesh refinement, whereas CR approaches a more visible accuracy ceiling.
\begin{table}[!t]
\caption{RMSE of Interpolation Methods Across Grid Resolutions}
\label{interp_rmse}
\centering
\scriptsize
\renewcommand{\arraystretch}{1.3}
\setlength{\tabcolsep}{3pt}
\begin{tabular}{
    >{\centering\arraybackslash}m{2cm} 
    >{\centering\arraybackslash}m{1.2cm} 
    >{\centering\arraybackslash}m{1.5cm} 
    >{\centering\arraybackslash}m{1.5cm} 
    >{\centering\arraybackslash}m{1.5cm}}
\toprule
Altitude Range (km) & Grid & GD (p=5) & GD (p=3) & CR \\
\midrule
\multirow{6}{*}{65--150}
  & V1 & $1.15$            & $0.88$             & $0.88$ \\
  & V2 & $4.25 \times 10^{-3}$ & $3.38 \times 10^{-3}$ & $3.47 \times 10^{-3}$ \\
  & V3 & $1.35 \times 10^{-3}$ & $4.68 \times 10^{-4}$ & $5.68 \times 10^{-4}$ \\
  & V4 & $4.25 \times 10^{-3}$ & $3.38 \times 10^{-3}$ & $3.48 \times 10^{-3}$ \\
  & V5 & $4.25 \times 10^{-3}$ & $3.38 \times 10^{-3}$ & $3.33 \times 10^{-3}$ \\
  & V6 & $4.25 \times 10^{-3}$ & $3.38 \times 10^{-3}$ & $3.47 \times 10^{-3}$ \\
\midrule
\multirow{6}{*}{150--200}
  & V1 & $5.62 \times 10^{-5}$ & $6.38 \times 10^{-5}$ & $9.19 \times 10^{-4}$ \\
  & V2 & $1.80 \times 10^{-5}$ & $2.00 \times 10^{-5}$ & $9.22 \times 10^{-4}$ \\
  & V3 & $1.80 \times 10^{-5}$ & $2.00 \times 10^{-5}$ & $9.22 \times 10^{-4}$ \\
  & V4 & $2.02 \times 10^{-5}$ & $1.81 \times 10^{-5}$ & $9.31 \times 10^{-4}$ \\
  & V5 & $1.53 \times 10^{-5}$ & $1.54 \times 10^{-5}$ & $2.88 \times 10^{-5}$ \\
  & V6 & $1.80 \times 10^{-5}$ & $2.00 \times 10^{-5}$ & $9.22 \times 10^{-4}$ \\
\midrule
\multirow{6}{*}{200--400}
  & V1 & $1.31 \times 10^{-3}$ & $1.36 \times 10^{-3}$ & $4.97 \times 10^{-3}$ \\
  & V2 & $1.97 \times 10^{-4}$ & $2.14 \times 10^{-4}$ & $4.78 \times 10^{-3}$ \\
  & V3 & $2.01 \times 10^{-4}$ & $2.19 \times 10^{-4}$ & $4.78 \times 10^{-3}$ \\
  & V4 & $1.19 \times 10^{-4}$ & $1.20 \times 10^{-4}$ & $4.81 \times 10^{-3}$ \\
  & V5 & $1.17 \times 10^{-4}$ & $1.18 \times 10^{-4}$ & $2.39 \times 10^{-4}$ \\
  & V6 & $1.93 \times 10^{-4}$ & $2.09 \times 10^{-4}$ & $4.78 \times 10^{-3}$ \\
\bottomrule
\end{tabular}

\vspace{1.5ex}
\begin{minipage}{\columnwidth}
\footnotesize
\textbf{Note:}Root-mean-square error (RMSE) values shown in the table are in units of electrons per cubic meter (el/m$^3$).\\
Grid definitions: V1: $10\,\mathrm{km} \times 1^\circ \times 1^\circ$,  
V2: $1\,\mathrm{km} \times 1^\circ \times 1^\circ$,  
V3: $0.1\,\mathrm{km} \times 1^\circ \times 1^\circ$,  
V4: $1\,\mathrm{km} \times 0.1^\circ \times 1^\circ$,  
V5: $1\,\mathrm{km} \times 0.01^\circ \times 1^\circ$,  
V6: $1\,\mathrm{km} \times 1^\circ \times 0.1^\circ$.  
Format: altitude $\times$ latitude $\times$ longitude.
\end{minipage}
\end{table}

Another notable feature is the directional sensitivity of the two schemes. 
GD responds more strongly to vertical refinement while remaining comparatively stable under horizontal refinement, whereas CR shows a more pronounced dependence on latitude resolution. 
This behavior is consistent with the fact that GD explicitly incorporates derivative information through its Hermite-type construction, which is advantageous in steep-gradient regions. 
Taken together, the results show that GD interpolation provides superior convergence behavior and directional robustness for ionospheric electron-density reconstruction, 
particularly under vertical refinement, and therefore offers more reliable support for subsequent 3D ray tracing and synthetic oblique-ionogram generation.

\subsection{3D Ray Tracing Performance Analysis}
To evaluate the performance of the proposed RTM-GD method, ray-tracing simulations are conducted 
under a range of representative HF propagation scenarios, 
complemented by synthetic oblique ionograms experiments based on ray-tracing results.

The three-dimensional electron density fields used in the ray-tracing experiments are generated using the IRI-2020 model, a widely adopted empirical ionospheric reference model, 
which serves as a standardized and reproducible benchmark environment. The proposed RTM-GD framework is model-independent 
and can be directly applied to arbitrary discrete electron density grids. The selected scenarios cover different solar activity levels, 
seasons, local times, geographic locations, operating frequencies, and both O- and X-mode propagation, ensuring representative HF ionospheric conditions 
for systematic performance evaluation.

\subsubsection{Performance Metrics}
\begin{figure*}[!t]
   \centering
   \subfloat[]{
       \includegraphics[width=0.30\textwidth]{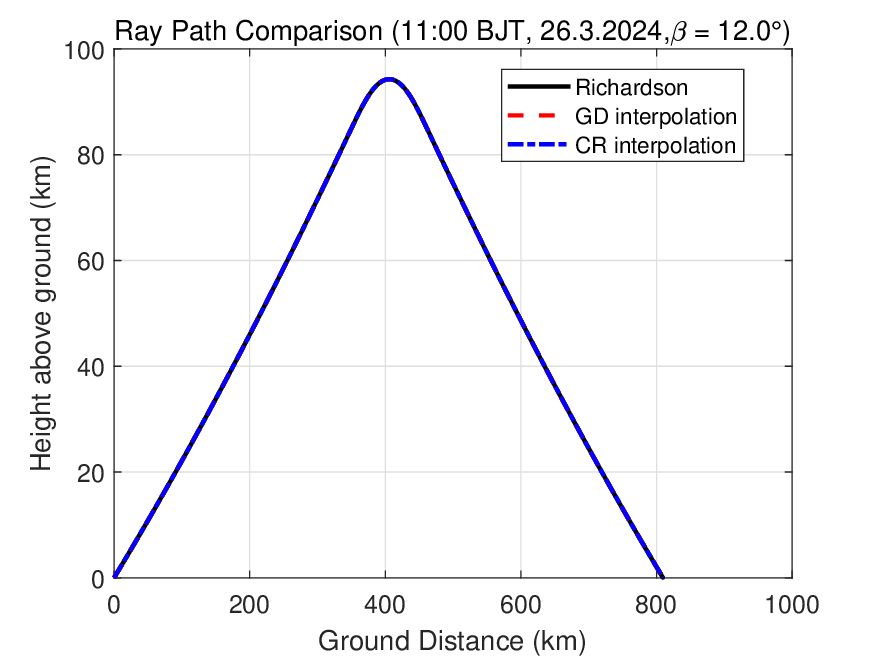}
       \label{fig:raypath_e}
   }
   \hfill
   \subfloat[]{
       \includegraphics[width=0.30\textwidth]{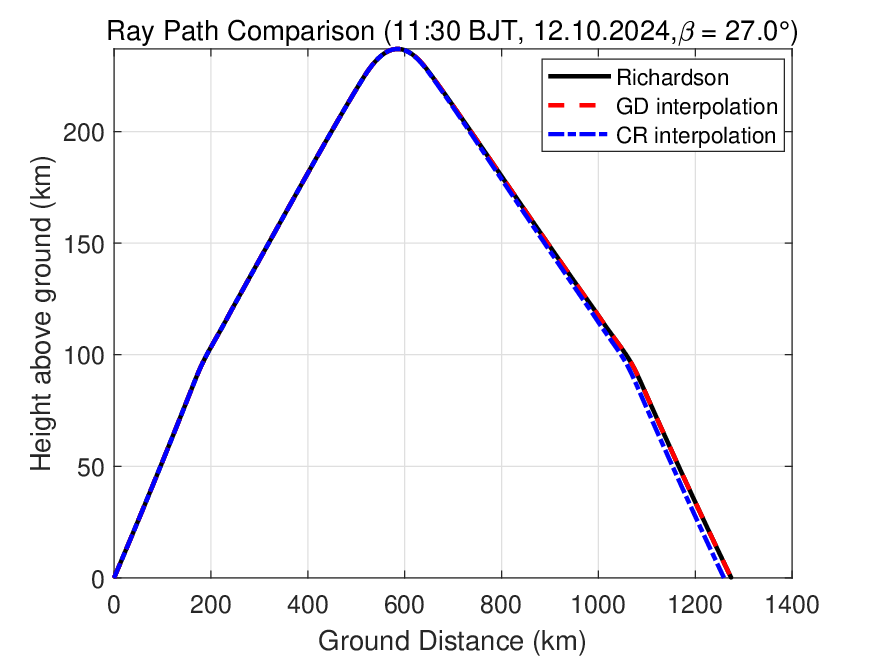}
       \label{fig:raypath_f}
   }
   \hfill
   \subfloat[]{
       \includegraphics[width=0.30\textwidth]{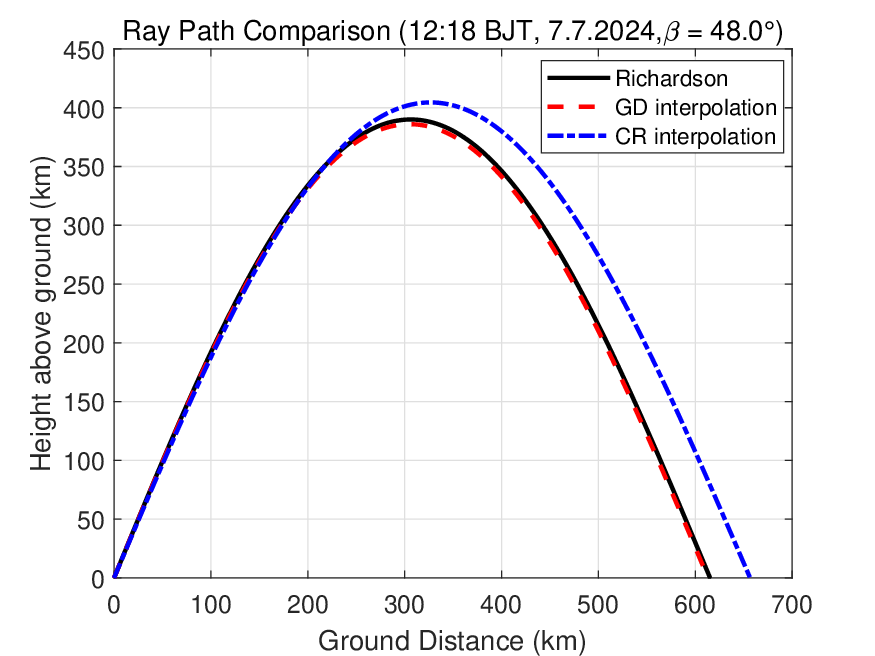}
       \label{fig:raypath_f2}
   }
   \caption{Ray-path comparison using Richardson extrapolation, GD interpolation, and CR interpolation for representative E-layer, F-layer, and high-elevation F2-mode HF propagation scenarios.}
   \label{fig:raypath_compare}
\end{figure*}

Representative ray-path plots in Fig.~\ref{fig:raypath_compare} are provided to qualitatively illustrate typical HF propagation scenarios 
and to verify the geometric consistency of different ray-tracing methods. 
Quantitative accuracy is then evaluated using the following physical error metrics.

\begin{subequations}\label{eq:rmse_metrics}
\begin{align}
\mathrm{GP}_{\mathrm{RMSE}} &= \sqrt{\frac{1}{N} \sum_{i=1}^N \left(\mathrm{GP}_{\text{interp},i} - \mathrm{GP}_{\text{ref},i}\right)^2}, \\
\mathrm{GD}_{\mathrm{RMSE}} &= \sqrt{\frac{1}{N} \sum_{i=1}^N \left(\mathrm{GD}_{\text{interp},i} - \mathrm{GD}_{\text{ref},i}\right)^2}, \\
\mathrm{AD}_{\mathrm{RMSE}} &= \sqrt{\frac{1}{N} \sum_{i=1}^N \left(\mathrm{AD}_{\text{interp},i} - \mathrm{AD}_{\text{ref},i}\right)^2}.
\end{align}
\end{subequations}
where \( N \) denotes the number of rays in each scan. 
Reference trajectories were generated using the high-precision Richardson extrapolation method with a step size of $\Delta h = 10^{-4}$ to ensure benchmark accuracy.

The accuracy in elevation-angle scanning experiments was evaluated using the root-mean-square errors (RMSE) of three parameters: 
group path (cumulative ionospheric delay), ground distance (horizontal ray projection at ground level), 
and azimuth deviation (angular shift from the launch direction). 
Computational efficiency was quantified by the total scanning time, $T_{\text{scan}}$.

In synthetic oblique ionogram experiments, qualitative evaluation is conducted by comparing the structural features of synthetic and measured ionograms.
To further enable quantitative evaluation, the mean relative error (MRE) of the group path is introduced to quantify the discrepancy between synthetic and measured oblique ionograms.
\begin{equation}
\mathrm{MRE} = \frac{1}{N} \sum_{k=1}^{N}
\frac{\left| P_{\mathrm{syn}}(f_k) - P_{\mathrm{raw}}(f_k) \right|}
{P_{\mathrm{raw}}(f_k)} \times 100\%,
\label{eq:mre_ionogram}
\end{equation}
where $P_{\mathrm{syn}}(f_k)$ and $P_{\mathrm{raw}}(f_k)$ denote the group path lengths of the synthetic and 
measured oblique ionograms at frequency $f_k$, respectively, and $N$ is the number of matched frequency points.

\subsubsection{Elevation-Angle Scanning Experiments}
To evaluate HF ray propagation performance across a broad range of realistic conditions, 
ray tracing simulations are performed under varying ionospheric states, 
elevation angle regimes (low and high), operating frequencies, and ordinary and extraordinary wave modes.

As a representative case, we simulate the scenario at 12:18 BJT on July 7, 2024, and 
benchmark RTM-GD against CR interpolation and a high-accuracy 
Richardson extrapolation reference. This time corresponds to typical mid-latitude 
summer daytime conditions with coexisting E and F layers, and is therefore adopted as 
a representative benchmark for the elevation-angle scanning experiments.

\begin{figure}[!t]
   \centering
   \subfloat[]{
       \includegraphics[width=0.35\textwidth]{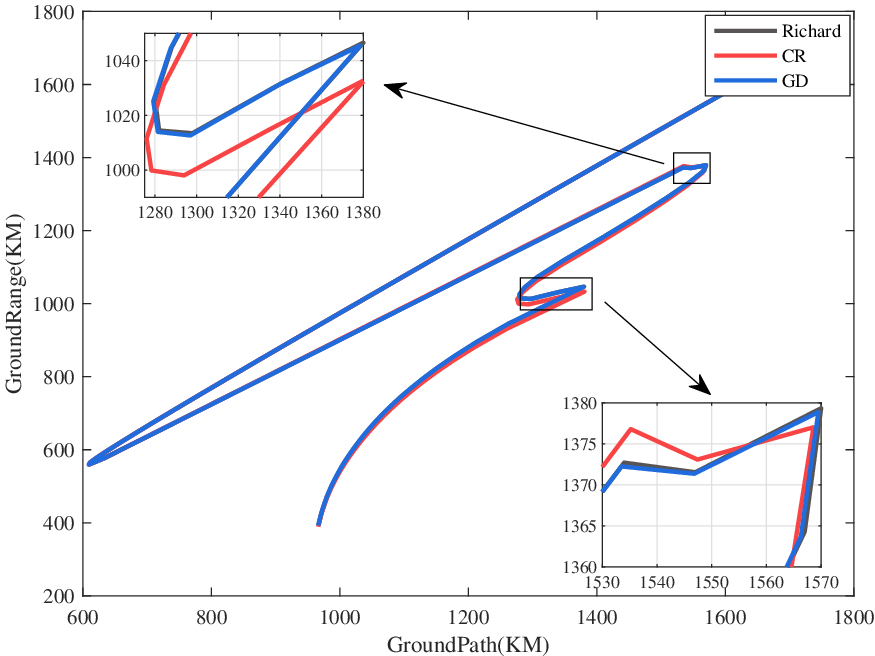}
       \label{fig:PD}
   }
   \hfill
   \subfloat[]{
       \includegraphics[width=0.35\textwidth]{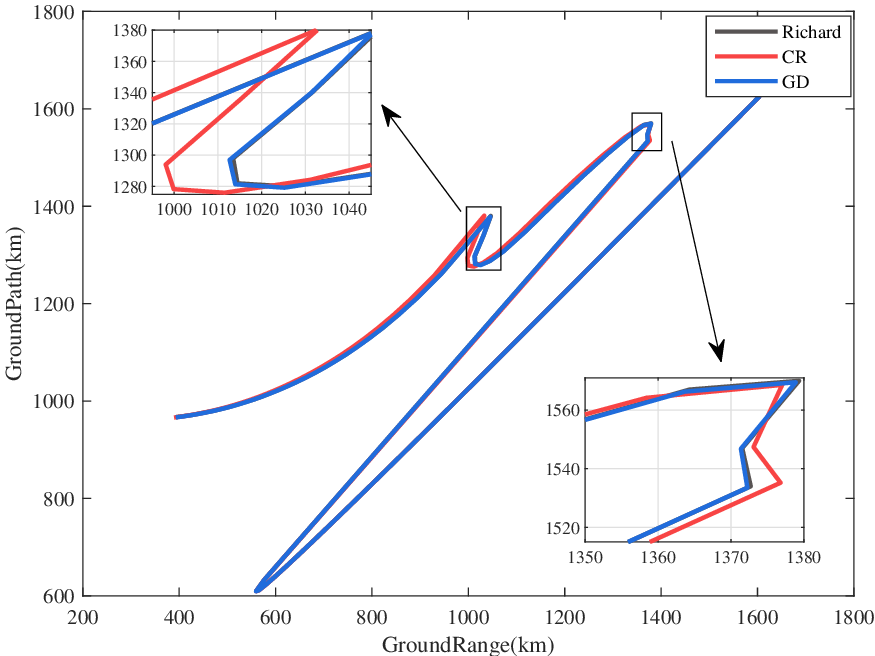}
       \label{fig:DP}
   }
   \caption{Comparison of interpolation methods for group path and ground distance reconstruction: 
   (a) Group path vs. ground distance (P–D); (b) Ground distance vs. group path (D–P).}
   \label{fig:gp_gd_rel}
\end{figure}

Figure~\ref{fig:gp_gd_rel} illustrates the relationship between group path and ground distance in both the forward and inverse representations. 
Compared with CR interpolation, RTM-GD follows the reference solution much more closely, not only in the global trend but also in the local geometric details. 
This agreement indicates that RTM-GD preserves the ray-path structure more faithfully over a broad elevation-angle range 
and provides more reliable trajectory reconstruction than the conventional interpolation scheme.

\begin{figure}[!t]
   \centering
   \includegraphics[width=0.95\linewidth]{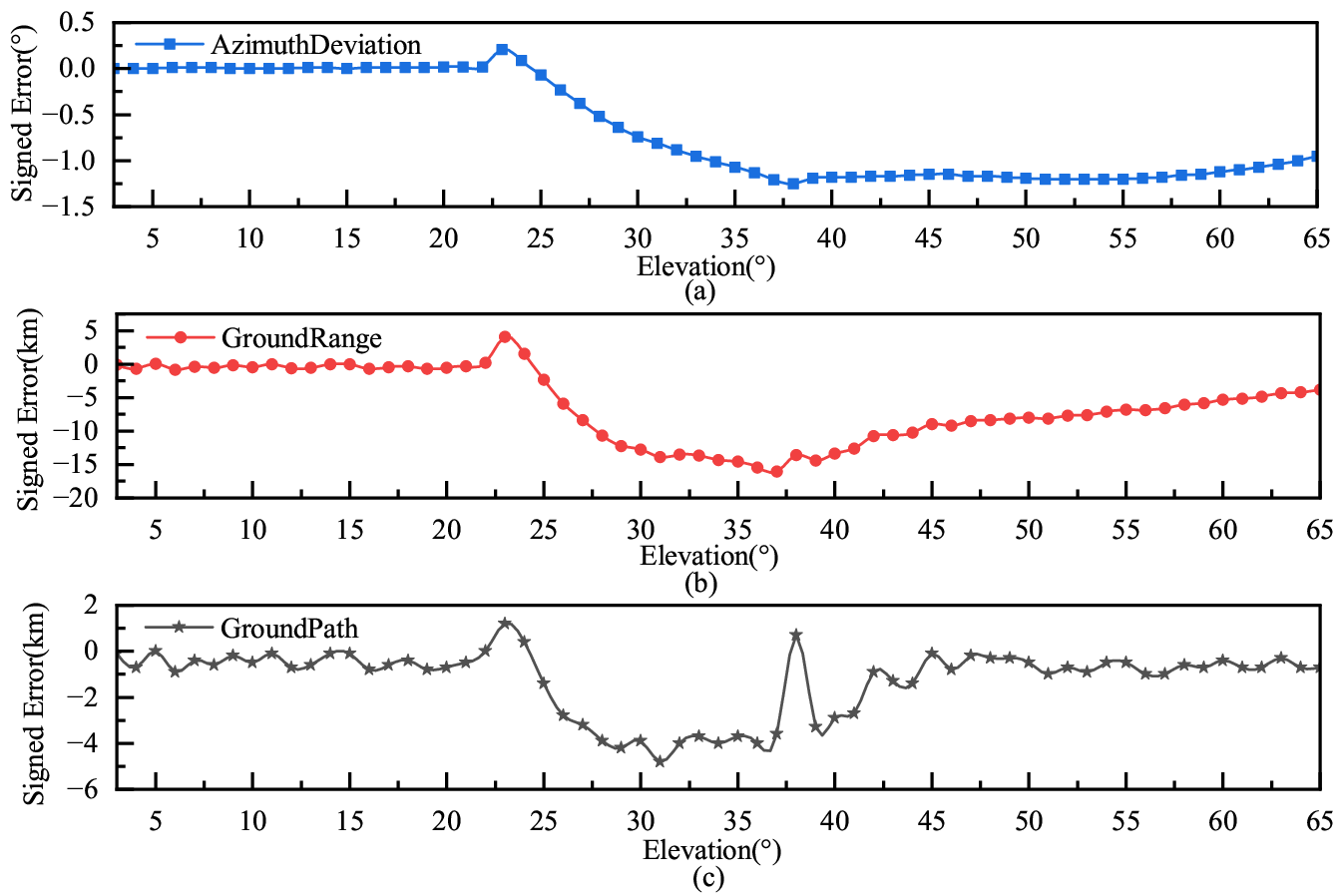}
   \caption{Ray-tracing errors of CR interpolation at 10\,MHz for O-mode propagation. Subplots show signed deviations in (a) azimuth, (b) ground distance, and (c) group path.}
   \label{fig:total_eleva_CR_error}
\end{figure}

\begin{figure}[!t]
   \centering
   \includegraphics[width=0.95\linewidth]{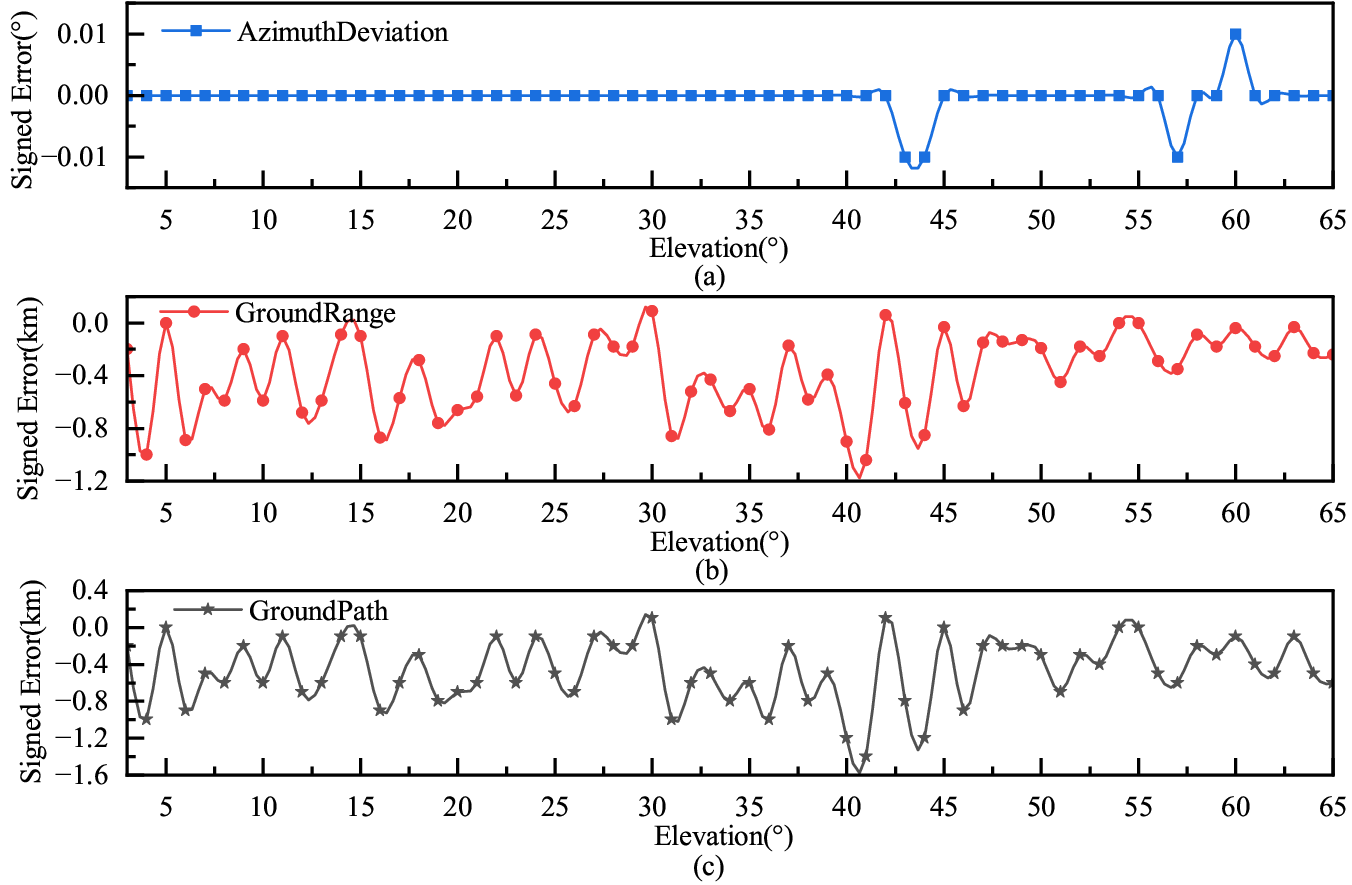}
   \caption{Ray-tracing errors of RTM-GD at 10\,MHz for O-mode propagation. Subplots show signed deviations in (a) azimuth, (b) ground distance, and (c) group path.}
   \label{fig:total_eleva_GD_error}
\end{figure}

Figure~\ref{fig:total_eleva_CR_error} further shows that the largest CR-related discrepancies occur in the elevation interval 
from $23^\circ$ to $37^\circ$, which corresponds to the E--F transition region. 
Within this range, the group-path error reaches $-4.8$~km and the ground-distance deviation increases to $-16.04$~km. 
These results suggest that CR interpolation is less capable of resolving path-parameter variations in the presence of steep vertical electron-density gradients.
In contrast, Fig.~\ref{fig:total_eleva_GD_error} shows that RTM-GD maintains consistent 
and stable accuracy throughout the scanning range, with group path errors within $[-1.4,0.1]$~km, 
97\% of ground distance errors within $\pm1$~km, and negligible azimuth deviations. 
This underscores the strong adaptability and numerical robustness of RTM-GD in steep gradient conditions.

Furthermore, Table~\ref{tab:iono_dependency_results} summarizes the ray-tracing accuracy and runtime performance of RTM-GD under a set of representative ionospheric scenarios, 
thereby providing a systematic evaluation of its robustness under diverse HF propagation conditions.
Richardson extrapolation serves as the high-precision reference but requires approximately one hour per elevation scan. 
Across all tested scenarios, RTM-GD maintains sub-kilometer RMSE in both group-path and ground-distance metrics 
and sub-0.01$^\circ$ azimuth deviation relative to Richardson extrapolation, while improving computational efficiency by 98\%. 
Compared to conventional CR interpolation, RTM-GD incurs less than 4\% additional computational cost yet reduces the RMSEs 
of ray parameters by approximately one order of magnitude for both propagation modes. 
Thus, RTM-GD demonstrates a favorable balance between accuracy and efficiency, 
making it suitable for 3D ray tracing under diverse ionospheric conditions.

\begin{table*}[htbp]
\caption{Ray Tracing Accuracy and Runtime Comparison under Diverse Ionospheric Conditions}
\label{tab:iono_dependency_results}
\centering
\scriptsize
\renewcommand{\arraystretch}{1.3}
\setlength{\tabcolsep}{4pt}
\begin{tabular}{m{0.9cm} m{2.6cm} m{2.8cm} m{1.1cm} m{0.9cm} m{0.9cm}
                m{1.4cm} m{1.4cm} m{1.4cm} m{1.3cm} m{1.6cm}}
\toprule
Exp. ID & Date / Beijing Time & Tx Location & Freq (MHz) & Wave & Method
& GP\_RMSE (km) & GD\_RMSE (km) & AD\_RMSE (deg) & Runtime (min) & Ref. Runtime (min) \\
\midrule
\multirow{2}{*}{D1} & \multirow{2}{*}{2019-01-15 / 02:00} & \multirow{2}{*}{(30$^\circ$N, 112$^\circ$E)} & \multirow{2}{*}{15} & \multirow{2}{*}{X}
& CR & 2.7706 & 3.0642 & 0.1906 & 1.03 & \multirow{2}{*}{87.51} \\
&  &  &  &  & GD & 0.3427 & 0.3144 & 0.0012 & 1.05 &  \\

\multirow{2}{*}{D2} & \multirow{2}{*}{2019-10-10 / 12:00} & \multirow{2}{*}{(32$^\circ$N, 118$^\circ$E)} & \multirow{2}{*}{10} & \multirow{2}{*}{O}
& CR & 18.5410 & 18.1260 & 0.2371 & 0.92 & \multirow{2}{*}{69.77} \\
&  &  &  &  & GD & 0.6932 & 0.6528 & 0.0049 & 0.94 &  \\

\multirow{2}{*}{D3} & \multirow{2}{*}{2019-04-05 / 09:00} & \multirow{2}{*}{(25$^\circ$N, 110$^\circ$E)} & \multirow{2}{*}{15} & \multirow{2}{*}{X}
&  CR & 0.8383 & 6.2835 & 1.1514 & 0.74 & \multirow{2}{*}{58.29} \\
&  &  &  &  & GD & 0.0926 & 0.0824 & 0.0001 & 0.76 &  \\

\multirow{2}{*}{D4} & \multirow{2}{*}{2022-11-16 / 14:00} & \multirow{2}{*}{(35$^\circ$N, 105$^\circ$E)} & \multirow{2}{*}{12} & \multirow{2}{*}{O}
& CR & 2.5127 & 8.6740 & 1.2629 & 0.84 & \multirow{2}{*}{74.17} \\
&  &  &  &  & GD & 0.8420 & 0.7282 & 0.0036 & 0.85 &  \\

\multirow{2}{*}{D5} & \multirow{2}{*}{2022-07-07 / 20:00} & \multirow{2}{*}{(40$^\circ$N, 120$^\circ$E)} & \multirow{2}{*}{8} & \multirow{2}{*}{X}
& CR & 2.2424 & 6.3467 & 0.4649 & 1.00 & \multirow{2}{*}{86.46} \\
&  &  &  &  & GD & 0.0309 & 0.0272 & 0.0001 & 1.04 &  \\

\multirow{2}{*}{D6} & \multirow{2}{*}{2022-09-20 / 06:30} & \multirow{2}{*}{(20$^\circ$N, 108$^\circ$E)} & \multirow{2}{*}{10} & \multirow{2}{*}{O}
& CR & 0.3202 & 2.0508 & 0.2707 & 0.79 & \multirow{2}{*}{63.56} \\
&  &  &  &  & GD & 0.0642 & 0.0595 & 0.0000 & 0.81 &  \\

\multirow{2}{*}{D7} & \multirow{2}{*}{2024-03-26 / 10:00} & \multirow{2}{*}{(27$^\circ$N, 112$^\circ$E)} & \multirow{2}{*}{15} & \multirow{2}{*}{X}
& CR & 0.6102 & 4.7331 & 0.6753 & 0.75 & \multirow{2}{*}{65.37} \\
&  &  &  &  & GD & 0.0873 & 0.0745 & 0.0012 & 0.78 &  \\

\multirow{2}{*}{D8} & \multirow{2}{*}{2024-06-18 / 16:00} & \multirow{2}{*}{(22$^\circ$N, 114$^\circ$E)} & \multirow{2}{*}{12} & \multirow{2}{*}{O}
& CR & 0.6064 & 4.2254 & 0.5952 & 0.89 & \multirow{2}{*}{74.50} \\
&  &  &  &  & GD & 0.0309 & 0.0293 & 0.0000 & 0.90 &  \\

\multirow{2}{*}{D9} & \multirow{2}{*}{2024-08-13 / 23:00} & \multirow{2}{*}{(45$^\circ$N, 130$^\circ$E)} & \multirow{2}{*}{8} & \multirow{2}{*}{X}
& CR & 0.3900 & 4.6688 & 0.9056 & 0.72 & \multirow{2}{*}{56.00} \\
&  &  &  &  & GD & 0.0356 & 0.0338 & 0.0001 & 0.73 &  \\

\multirow{2}{*}{D10} & \multirow{2}{*}{2024-12-10 / 11:30} & \multirow{2}{*}{(30$^\circ$N, 135$^\circ$E)} & \multirow{2}{*}{10} & \multirow{2}{*}{O}
& CR & 2.1302 & 7.6482 & 1.0979 & 0.94 & \multirow{2}{*}{78.79} \\
&  &  &  &  & GD & 0.1315 & 0.0650 & 0.0002 & 0.95 &  \\
\bottomrule
\end{tabular}
\vspace{1.5ex}
\begin{minipage}{\textwidth}
\footnotesize \textit{Note:} ``Runtime'' denotes the wall-clock time for one elevation-angle scan, while ``Reference Runtime'' refers to the computation time using Richardson extrapolation.
\end{minipage}
\end{table*}

\subsubsection{High-Elevation F2-Mode Scenarios}
This investigation examines high-elevation propagation within the F2 layer, a regime characterized by strong refractive curvature 
and heightened numerical sensitivity. To generate propagation paths approaching critical incidence, 
a transmission frequency of 15 MHz was selected, and elevation angles were scanned from $39.2^\circ$ to $48.0^\circ$. 
This range spans the main high-elevation F2-mode propagation region ($39.2^\circ$–$47.8^\circ$) and extends into 
the near-critical zone ($47.8^\circ$–$48.0^\circ$), where ray trajectories approach a horizontal turning point. 
The numerical robustness of RTM-GD was assessed by statistically analyzing the root-mean-square errors (RMSEs) 
of key path parameters over this angular interval.

\begin{figure}[!t]
   \centering
   \includegraphics[width=0.95\linewidth]{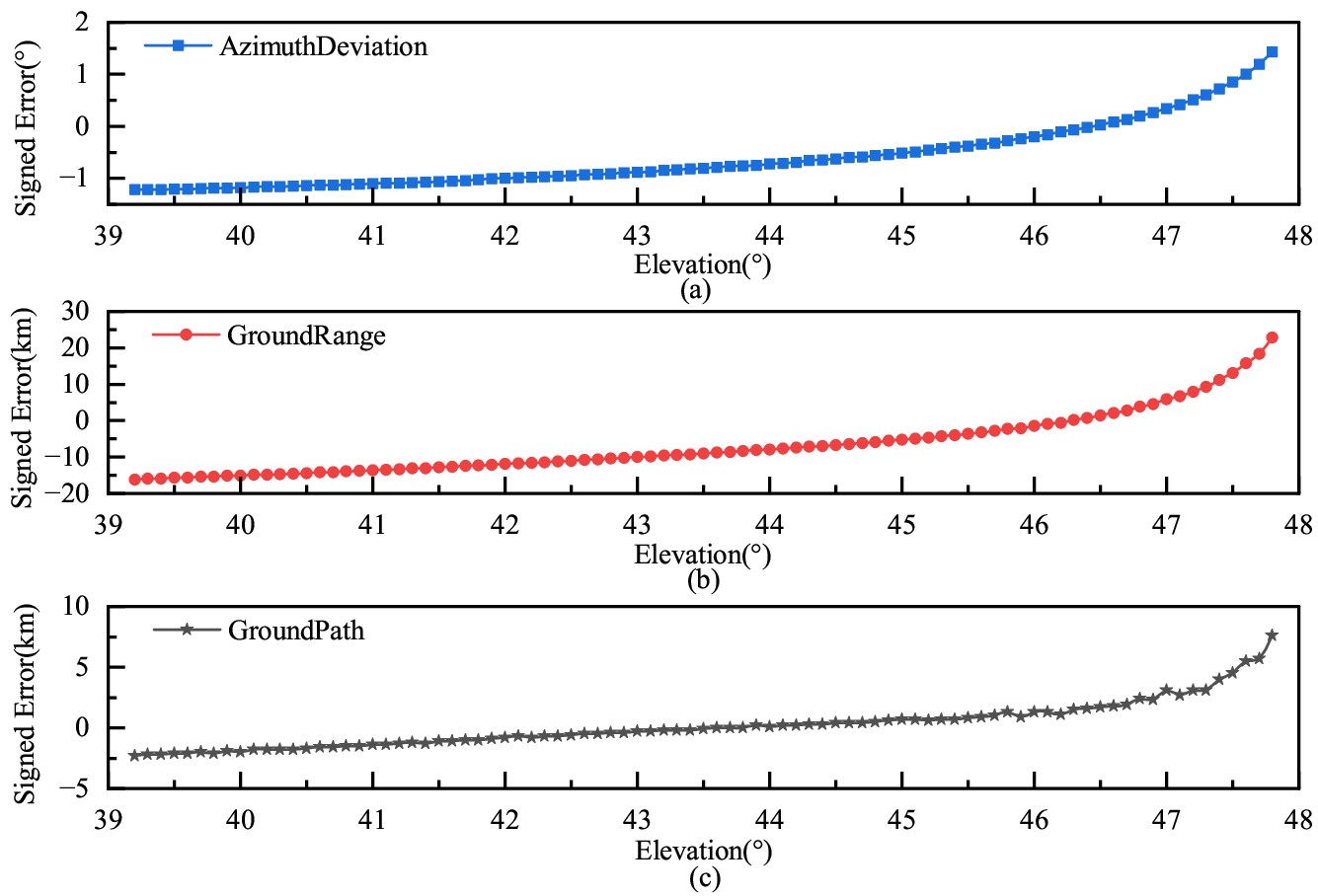}
   \caption{Ray-tracing errors of CR interpolation across elevation angles from 39.2$^\circ$ to 47.8$^\circ$ for O-mode propagation at 15\,MHz. 
   Subplots show deviations in (a) azimuth, (b) ground distance, and (c) group path.}
   \label{fig:cr_main}
\end{figure}

\begin{figure}[!t]
   \centering
   \includegraphics[width=0.95\linewidth]{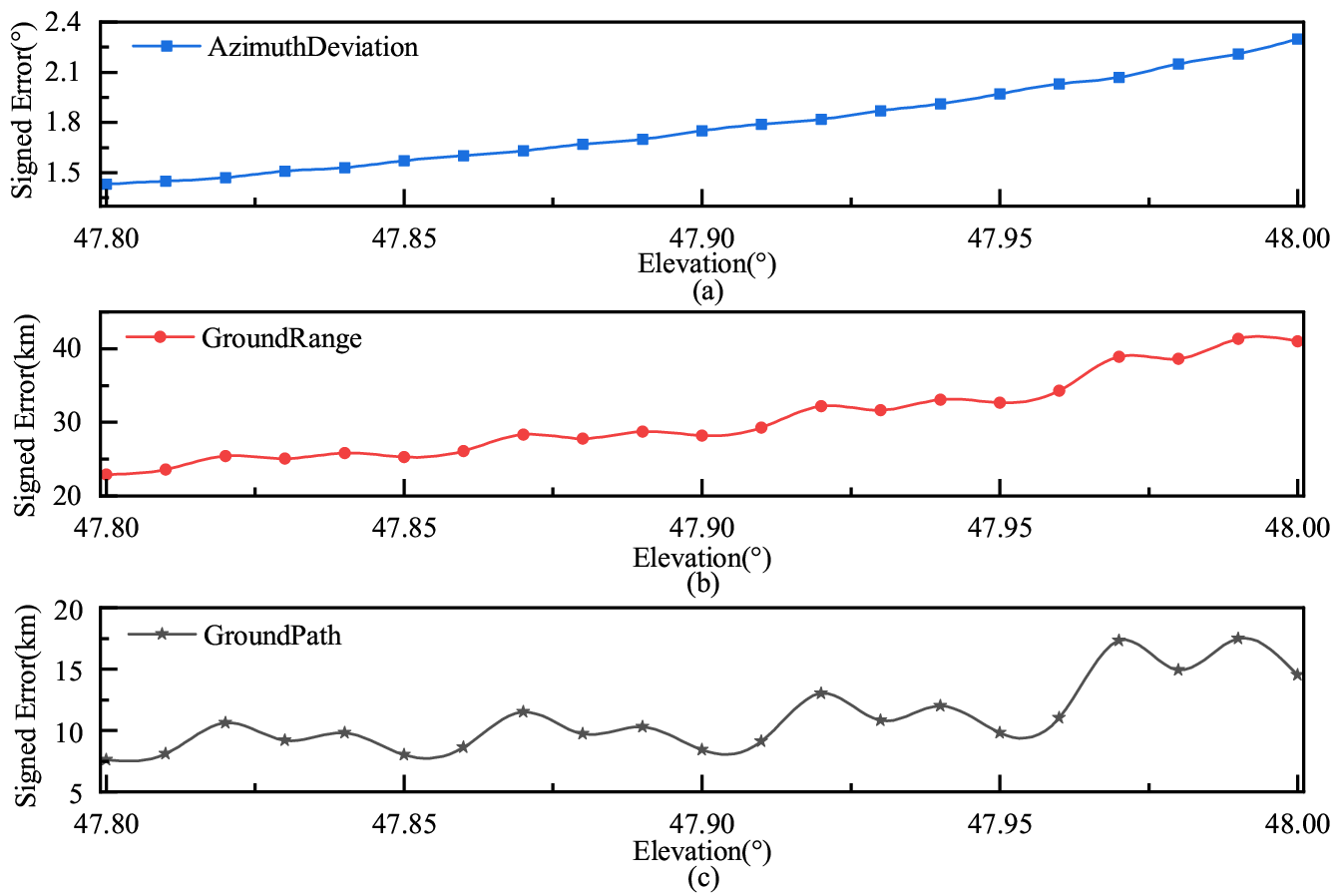}
   \caption{Ray-tracing errors of CR interpolation across elevation angles from 47.8$^\circ$ to 48$^\circ$ for O-mode propagation at 15\,MHz. 
   Subplots show deviations in (a) azimuth, (b) ground distance, and (c) group path.}
   \label{fig:cr_transition}
\end{figure}

Figs.~\ref{fig:cr_main} and \ref{fig:cr_transition} show that CR interpolation produces pronounced deviations in the high-elevation F2-mode regime. 
Over the main propagation interval, the group-path error rises to $7.6$ km, 
the ground-distance error reaches $22.9$ km, and the azimuth deviation is about $1.43^\circ$. 
The discrepancy becomes even more severe as the launch angle moves toward the critical-incidence region, 
where the corresponding errors increase to $14.5$ km in group path, $41$ km in ground distance, and $2.3^\circ$ in azimuth.

The rapid growth of the error near critical incidence can be attributed to the combined influence of 
a nearly horizontal turning point and the strong sensitivity of the ray solution to higher-order variations in the steep vertical electron-density gradient. 
Under these conditions, even small interpolation-induced perturbations in the refractive-index gradient may accumulate noticeably during numerical integration. 
Because CR interpolation does not provide sufficiently smooth and physically consistent gradient information in this regime, 
the resulting trajectory is more prone to drift and numerical instability.

\begin{figure}[!t]
   \centering
   \includegraphics[width=0.95\linewidth]{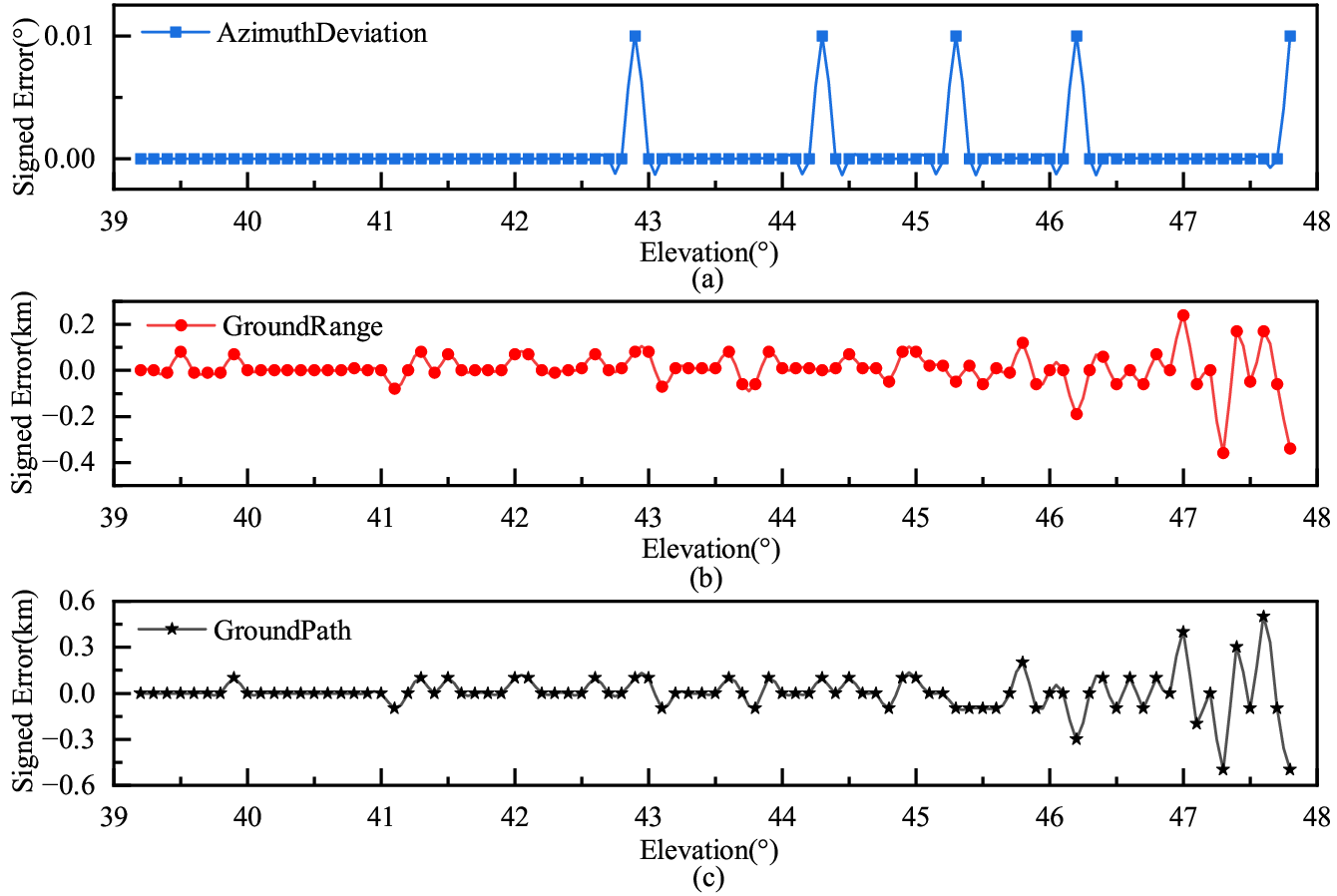}
   \caption{Ray-tracing errors of RTM-GD across elevation angles from 39.2$^\circ$ to 47.8$^\circ$ for O-mode propagation at 15\,MHz. 
   Subplots show deviations in (a) azimuth, (b) ground distance, and (c) group path.}
   \label{fig:gd_main}
\end{figure}

\begin{figure}[!t]
   \centering
   \includegraphics[width=0.95\linewidth]{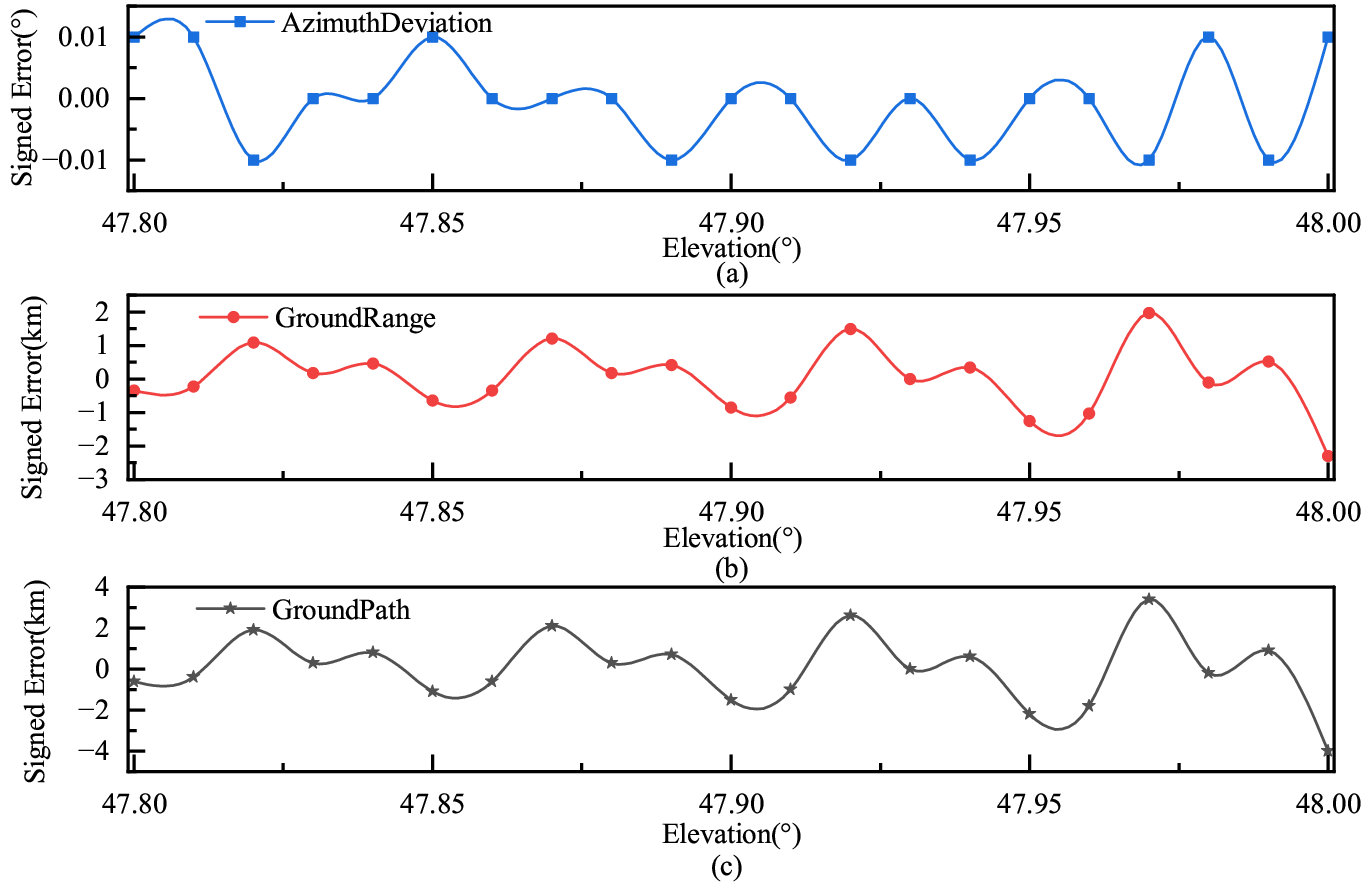}
   \caption{Ray-tracing errors of RTM-GD across elevation angles from 47.8$^\circ$ to 48$^\circ$ for O-mode propagation at 15\,MHz. 
   Subplots show deviations in (a) azimuth, (b) ground distance, and (c) group path.}
   \label{fig:gd_transition}
\end{figure}

By comparison, RTM-GD exhibits substantially better behavior under the same propagation conditions, as shown in Figs.~\ref{fig:gd_main} and \ref{fig:gd_transition}. 
Across the main propagation interval, the group-path error stays below $0.5$ km, the ground-distance deviation remains within $0.36$ km, and the azimuth error is essentially negligible. 
Even when the ray approaches the near-critical region, RTM-GD continues to maintain strong error control, with group-path and ground-distance errors confined to $4$ km and $2.3$ km, 
respectively, while the azimuth deviation remains below $0.01^\circ$ and no evident numerical fluctuation is observed.

These results demonstrate that RTM-GD maintains high accuracy and numerical stability under 
demanding high-elevation F2-mode propagation conditions. This robustness is attributed to 
the $C^1$-continuous electron density reconstruction provided by GD interpolation, 
which ensures smooth refractive index gradients and supports stable Hamiltonian ray integration near critical incidence.

\subsubsection{Synthesis of Oblique Ionograms}
To evaluate the performance of the proposed RTM-GD ray-tracing method under realistic HF propagation conditions, 
experiments on synthetic oblique ionograms are conducted. Two representative HF propagation links are selected, 
namely the Suzhou-Wuhan link and the Lanzhou-Wuhan link. For each link, multiple representative moments 
are chosen to construct synthetic oblique ionograms, so as to reflect different ionospheric propagation conditions.

The three-dimensional electron density fields are generated using the IRI-2020 model and then continuously reconstructed 
via GD interpolation. Based on the reconstructed electron density fields, ray tracing is performed through 
a joint frequency-elevation scanning strategy, yielding the group-path distributions of the synthetic oblique ionograms.

\begin{figure*}[!t]
    \centering
    \includegraphics[width=0.95\textwidth]{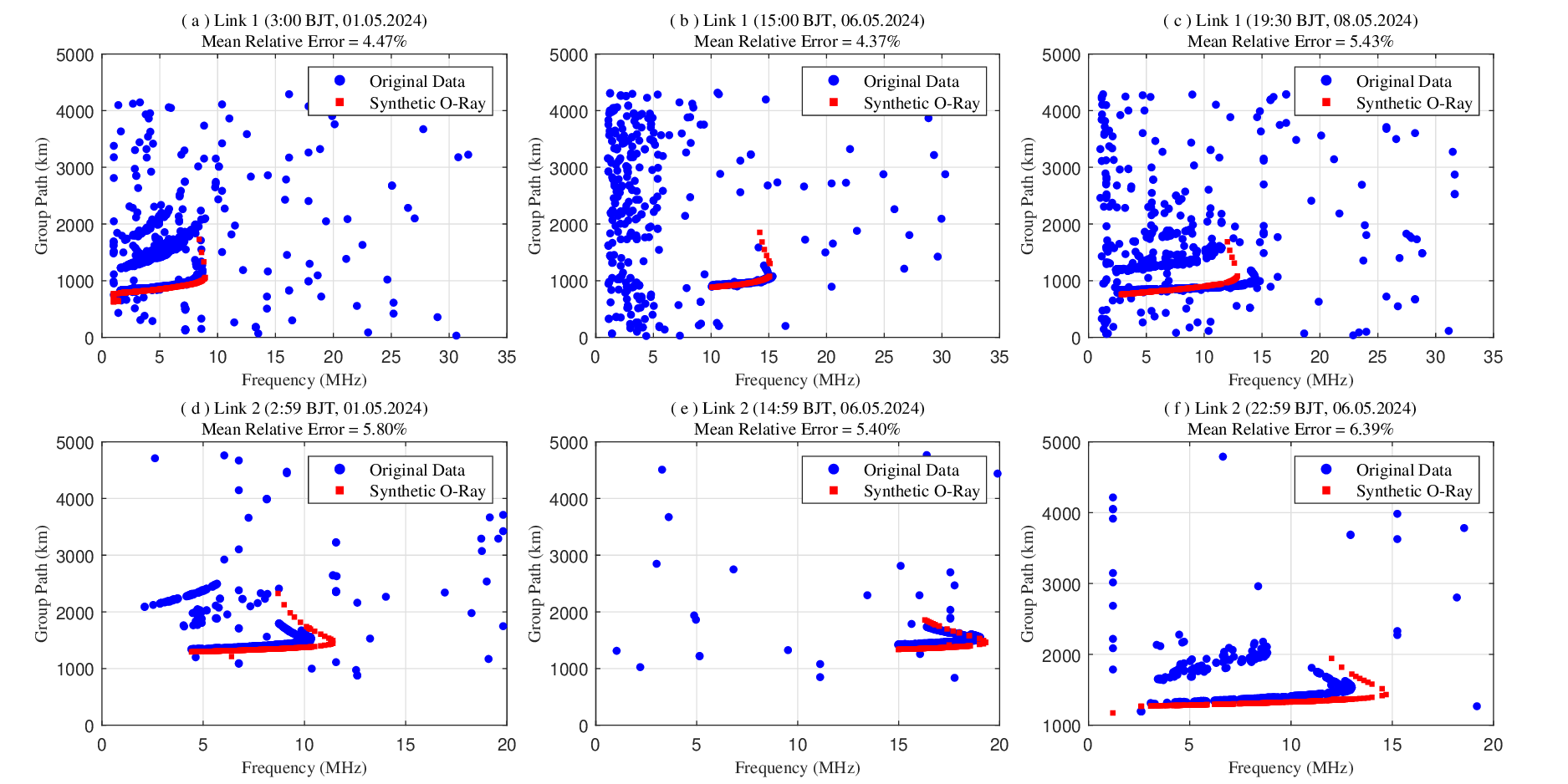}
    \caption{Comparison between measured and synthetic oblique ionograms for two representative HF propagation links.}
    \label{fig:IonogSyn}
\end{figure*}

As shown in Figure~\ref{fig:IonogSyn}, the synthetic oblique ionograms produced by the RTM-GD method exhibit clear and continuous echo traces. 
Both the overall trends and local features of the synthetic ionograms are in close agreement with the corresponding measured ionograms, 
successfully reproducing the typical low-elevation and high-elevation F2-mode echo structures.
The consistency between the synthetic and measured ionograms is further quantified using the mean relative error of the group path. 
Across all examined cases, this error remains within approximately 4.4\%--6.4\%, 
indicating good quantitative agreement between the synthetic and measured results.

Overall, the above results confirm that the ray paths generated by the RTM-GD method can accurately 
capture the actual physical propagation behavior of HF waves in the ionosphere. 
The good agreement in group-path parameters between the synthetic and measured oblique ionograms 
verifies the physical consistency and numerical reliability of the proposed ray-tracing approach.
Further improvements in the accuracy of synthetic ionograms may be achieved in future work 
by incorporating ionospheric parameter correction techniques.

\section{Conclusion}
\label{sec:conclusion}
This paper addressed the accuracy efficiency trade-off in HF ionospheric ray tracing under discretized electron density models. 
A ray-tracing framework combining Hamiltonian ray equations with a $C^1$-continuous Galerkin--Difference interpolation strategy was proposed, 
enabling the construction of an everywhere differentiable electron density field to support stable numerical integration and accurate propagation-path computation.

The proposed RTM-GD framework was validated through theoretical analysis and numerical experiments using both simulated and measured HF data. 
The results demonstrate that incorporating $C^1$-continuous electron density reconstruction into the ray-tracing process 
effectively improves numerical stability and achieves a favorable balance between propagation accuracy and computational efficiency 
across diverse ionospheric conditions.

Future work will extend the RTM-GD framework to time-varying ionospheric environments and data-assimilative models, 
and explore its application to inverse propagation and HF localization problems. 
These extensions are expected to further enhance the applicability of the proposed method to practical HF sensing, communication, 
and over-the-horizon radar systems.

\appendices
\section{\texorpdfstring{Basis Functions for $p=3$ Galerkin-Difference Interpolation}{Basis Functions for p=3 GD Interpolation}}
\label{appendix:gd_p3}
For $p=3$, the GD interpolation constructs a degree-5 piecewise polynomial over each interval, using a five-point stencil ($q = 2$). This appendix derives the explicit central difference weights and Hermite basis functions used to construct the 1D interpolation basis.

The first derivative operator \( D^{(1,3)} \) is approximated using a symmetric three-point central stencil with accuracy order \( \mathcal{O}(h^2) \). This compact formulation minimizes numerical dispersion and ensures $C^1$ continuity across adjacent cells:
\begin{equation}
h D^{(1,3)} u_j = \eta_{-1}^{(1,3)} u_{j-1} + \eta_0^{(1,3)} u_j + \eta_{1}^{(1,3)} u_{j+1},
\end{equation}
with weights given by:
\begin{equation}
\eta_{-1}^{(1,3)} = -\frac{1}{2}, \quad \eta_0^{(1,3)} = 0, \quad \eta_{1}^{(1,3)} = \frac{1}{2}, \quad \eta_{\pm 2}^{(1,3)} = 0.
\end{equation}

The Hermite basis functions satisfy:
\begin{linenomath*}
\begin{align}
H_{\alpha,0}^{(3,1)}(v) &= \delta_{\alpha,v}, \quad \alpha,v \in \{-2,\dots,3\} \\
\left.\frac{d}{dx} H_{\alpha,0}^{(3,1)}(v)\right|_{v=0,1} &= 0 , \quad \alpha \in \{-2,\dots,3\} \\
\left.\frac{d}{dx} H_{\alpha,1}^{(3,1)}(v)\right|_{v=0,1} &= \delta_{\alpha,v}, \quad \alpha \in \{0,1\} \\
H_{\alpha,1}^{(3,1)}(v) &= 0, \quad \alpha \in \{0,1\}, v \in \{-2,\dots,3\}
\end{align}
\end{linenomath*}

The global GD basis function \( \psi_k^{(3,1)}(x) \), as defined in Eq.~\eqref{eq:psi_case1} of the main text, combines Hermite polynomials and central difference weights over the subinterval \( \xi \in [k, k+1] \):
\begin{equation}
\begin{split}
\psi_k^{(3,1)}(x) = {} & H_{-k,0}^{(3,1)}(\xi - k) \\
& + \eta_{-k}^{(1,3)} H_{0,1}^{(3,1)}(\xi - k) \\
& + \eta_{-k-1}^{(1,3)} H_{1,1}^{(3,1)}(\xi - k).
\end{split}
\end{equation}
Explicit expressions for \( k = -2, -1, 0, 1 \) are:
\begin{subequations}
\begin{align}
\psi_{-2}^{(3,1)}(\xi) ={}& -\tfrac{1}{6}(\xi + 2)^5 + \tfrac{5}{12}(\xi + 2)^4 \notag\\
& + \tfrac{1}{6}(\xi + 2)^3 - \tfrac{5}{12}(\xi + 2)^2,\\
\psi_{-1}^{(3,1)}(\xi) ={}& \tfrac{1}{2}(\xi + 1)^5 - \tfrac{5}{4}(\xi + 1)^4 \notag\\
& - \tfrac{1}{2}(\xi + 1)^3 + \tfrac{7}{4}(\xi + 1)^2  + \tfrac{1}{2}(\xi + 1),\\
\psi_{0}^{(3,1)}(\xi) ={}& -\tfrac{1}{2}\xi^5 + \tfrac{5}{4}\xi^4 \notag\\
& + \tfrac{1}{2}\xi^3 - \tfrac{9}{4}\xi^2 + 1,\\
\psi_{1}^{(3,1)}(\xi) ={}& \tfrac{1}{6}(\xi - 1)^5 - \tfrac{5}{12}(\xi - 1)^4 \notag\\
& - \tfrac{1}{6}(\xi - 1)^3 + \tfrac{11}{12}(\xi - 1)^2  - \tfrac{1}{2}(\xi - 1).
\end{align}
\end{subequations}
Each function is compactly supported over its respective subinterval \( [k, k+1] \), vanishing elsewhere. Their smoothness and local support make them well-suited for efficient numerical interpolation on structured grids.

\section*{Acknowledgment}
The electron density and geomagnetic field data used in this study were obtained from the IRI-2020~\cite{bilitza2017international} and IGRF-13~\cite{alken2022special} models, respectively. The measured oblique ionograms used for experimental validation were kindly provided by Prof. Zhigang Zhang. The authors also thank Prof. Qing Xia for her insightful guidance on the mathematical formulation and theoretical analysis.

\bibliographystyle{IEEEtran}
\bibliography{references}

\end{document}